\documentstyle[12pt]{article}

\input epsf.tex

\newcommand{\tr}{\mbox{tr}}
\newcommand{\eqn}[1]{(\ref{#1})}
\newcommand{\del}{\partial}

\newcommand{\real}{{\bb R}} 

\def\(#1{ ^{(#1)} }

\font\mybb=msbm10 at 12pt
\def\bb#1{\hbox{\mybb#1}}

\def\e{{\rm e}}

\def\beq{\begin{equation}}
\def\eeq{\end{equation}}
\def\be{\begin{equation}}
\def\ee{\end{equation}}
\def\bea{\begin{eqnarray}}
\def\eea{\end{eqnarray}}
\def\bd{\begin{displaymath}}
\def\ed{\end{displaymath}}

\makeatletter
\newdimen\normalarrayskip              
\newdimen\minarrayskip                 
\normalarrayskip\baselineskip
\minarrayskip\jot
\newif\ifold             \oldtrue            \def\new{\oldfalse}
\def\arraymode{\ifold\relax\else\displaystyle\fi} 
\def\@arrayskip{\ifold\baselineskip\z@\lineskip\z@
     \else
     \baselineskip\minarrayskip\lineskip2\minarrayskip\fi}
\def\@arrayclassz{\ifcase \@lastchclass \@acolampacol \or
\@ampacol \or \or \or \@addamp \or
   \@acolampacol \or \@firstampfalse \@acol \fi
\edef\@preamble{\@preamble
  \ifcase \@chnum
     \hfil$\relax\arraymode\@sharp$\hfil
     \or $\relax\arraymode\@sharp$\hfil
     \or \hfil$\relax\arraymode\@sharp$\fi}}
\def\@array[#1]#2{\setbox\@arstrutbox=\hbox{\vrule
     height\arraystretch \ht\strutbox
     depth\arraystretch \dp\strutbox
     width\z@}\@mkpream{#2}\edef\@preamble{\halign \noexpand\@halignto
\bgroup \tabskip\z@ \@arstrut \@preamble \tabskip\z@ \cr}%
\let\@startpbox\@@startpbox \let\@endpbox\@@endpbox
  \if #1t\vtop \else \if#1b\vbox \else \vcenter \fi\fi
  \bgroup \let\par\relax
  \let\@sharp##\let\protect\relax
  \@arrayskip\@preamble}
\makeatother

\newcommand{\newsection}[1]
{\vspace{5mm}
\pagebreak[3]
\addtocounter{section}{1}
\setcounter{equation}{0}
\setcounter{subsection}{0}
\begin{flushleft}
{\large\bf \thesection. #1}
\end{flushleft}
\nopagebreak
\medskip
\nopagebreak}

\newlength{\extraspace}
\newlength{\extraspaces}
\begin{document}

\renewcommand{\footnotesize}{\small}

\addtolength{\baselineskip}{.8mm}

\thispagestyle{empty}

\begin{flushright}
\baselineskip=12pt
DSF/50-97\\
OUTP-97-61P\\
hep-th/9711012\\
\hfill{  }\\ November 1997
\end{flushright}
\vspace{.5cm}

\begin{center}

\baselineskip=12pt

{\Large\bf{Matrix $\sigma$-models for Multi D-brane Dynamics
}}\\[15mm]

{\sc Fedele Lizzi}\footnote{E-mail: {\tt lizzi@na.infn.it}}
\\[4mm]
{\it Dipartimento di Scienze Fisiche, Universit\`a di Napoli Federico II\\ and
INFN, Sezione di Napoli, Italy}
\\[8mm]
{\sc Nick E. Mavromatos\footnote{PPARC Advanced Fellow (U.K.).\\ E-mail: {\tt
n.mavromatos1@physics.oxford.ac.uk}}} {\sc and Richard J.\ Szabo\footnote{Work
supported in part by PPARC (U.K.).\\ E-mail: {\tt
r.szabo1@physics.oxford.ac.uk}}}
\\[4mm]
{\it Department of Physics -- Theoretical Physics, University of Oxford\\ 1
Keble Road, Oxford OX1 3NP, U.K.}
\\[15mm]

\vskip 1.0 in

{\sc Abstract}

\begin{center}
\begin{minipage}{15cm}

We describe a dynamical worldsheet origin for the Lagrangian describing the
low-energy dynamics of a system of parallel D-branes. We show how matrix-valued
collective coordinate fields for the D-branes naturally arise as couplings of a
worldsheet $\sigma$-model, and that the quantum dynamics require that these
couplings be mutually noncommutative. We show that the low-energy effective
action for the $\sigma$-model couplings describes the propagation of an open
string in the background of the multiple D-brane configuration, in which all
string interactions between the constituent branes are integrated out and the
genus expansion is taken into account, with a matrix-valued coupling. The
effective field theory is governed by the non-abelian Born-Infeld target space
action which leads to the standard one for D-brane field theory.

\end{minipage}
\end{center}

\end{center}

\noindent

\vfill
\newpage
\pagestyle{plain}
\setcounter{page}{1}
\setcounter{footnote}{0}
\stepcounter{subsection}

\newsection{Introduction}

With the advent of Dirichlet-brane field theory, introduced by Witten
\cite{witten1} and elucidated on by Taylor \cite{taylor}, there has been a lot
of activity in describing the short-distance spacetime structure of string
theory (see for example \cite{dbrane}). The D-brane action is obtained from
dimensional reduction, to the world-volume of the D-brane, of ten-dimensional
supersymmetric $U(N)$ Yang-Mills theory which describes the low-energy dynamics
of open superstrings with the usual Neumann boundary conditions. The target
space coordinate fields are now $N\times N$ matrices and hence describe some
sort of noncommutative spacetime geometry, giving an explicit realization of
the idea that at small distances the conventional notion of spacetime must be
abandoned.

The supersymmetric $N\times N$ matrix quantum mechanics for D-particles with
action
\beq
S_M=\frac1{4\pi\alpha'}\int
dt~\tr\left(\left(D_tY^i\right)^2-\frac{g_s}2\left[Y^i,Y^j\right]^2
+2\psi^{\rm T}D_t\psi-2\sqrt{g_s}\,\psi^{\rm
T}\gamma_i\left[\psi,Y^i\right]\right)
\label{Matrixaction}\eeq
has been of particular interest recently, in light of the conjecture
\cite{bfss} that in the large-$N$ limit it provides a Hamiltonian formalism for
the low-energy dynamics of M Theory in the infinite momentum frame of the
11-dimensional spacetime. The action \eqn{Matrixaction} is the reduction of
10-dimensional supersymmetric Yang-Mills theory down to (0 + 1)-dimensions,
with $D_t=\partial_t-i[A_0,\,\cdot\,]$. The fields $Y^i(t)$, $i=1,\dots,9$, are
$N\times N$ Hermitian matrices in the adjoint representation of $U(N)$
describing the collective coordinates of a system of $N$ parallel D0-branes
(with infinitesimal separation), and $\psi$ are their superpartners. The trace
in \eqn{Matrixaction} is taken in the fundamental representation of $U(N)$,
$1/2\pi\alpha'$ is the string tension, and $g_s$ is the (closed) string
coupling constant. The $N$ D-particles play the role of partons in the
light-cone theory and the large-$N$ limit naturally incorporates the quantum
field theoretical Fock space into the model. Among other things, the gauge
symmetry group $U(N)$ contains the permutation symmetry $S_N$ of $N$ identical
particles as its Weyl subgroup, and so the action \eqn{Matrixaction}
naturally describes a second quantized theory from the onset.

The D-brane field theory is canonically obtained from the target space
low-energy dynamics of open superstrings, and the string coupling $g_s$
originates from the ten-dimensional Yang-Mills coupling constant. In this
letter we will describe how to derive \eqn{Matrixaction} from a worldsheet
$\sigma$-model for a system of $N$ D-branes interacting via the exchange of
fundamental strings. We show how the matrix-valued D-brane configuration fields
$Y^i$ arise as couplings in such a $\sigma$-model. Starting with a
configuration where the D-branes lie on top of each other, we demonstrate that
the spacetime coordinates are necessarily described by matrices. We then
demonstrate that the quantum dynamics of the $N$ D-brane system imply that the
D-branes must acquire an infinitesimal separation.  The topological expansion
of the $\sigma$-model induces independent off-diagonal degrees of freedom
describing this separation, and it also provides a geometrical and dynamical
origin for the string coupling constant in the low-energy effective target
space action. This latter action, which is shown to be equivalent to
\eqn{Matrixaction} in the strong tension limit, describes the propagation (in
$\sigma$-model coupling constant space) of a single fundamental string in the
background of a ``fat brane", i.e. the multi D-brane system with all string
interactions between the constituent branes integrated out and with a
matrix-valued coupling to the open string. These results exhibit a worldsheet
origin for the appearance of matrices as coordinate fields at short distance
scales, and it may have implications for the recent attempts \cite{Mgeom,2b} at
a geometrical interpretation of \eqn{Matrixaction} as a sum over random
surfaces
appropriate to the perturbation expansion of string theory.

\newsection{Worldsheet $\sigma$-models and the Emergence of Matrices}

We start by showing how a worldsheet $\sigma$-model construction yields a
natural realization of one of the central ideas implied by Matrix Theory
\cite{witten1,bfss} -- that the spacetime induced by coincident D-branes is
described by noncommuting matrix-valued coordinates. Let us first describe the
action for a string in the background of a single D0-brane within the
$\sigma$-model approach\footnote{In this letter we shall, for definiteness,
discuss the case of D-particles, but the analysis and results trivially
generalize to D-instantons and extended D-branes as well.}. The open string is
described by a worldsheet $\Sigma$ which at tree-level we usually take to have
the topology of a disc. The endpoints of the string are fixed at the same point
in spacetime, the coordinates of the D0-brane. The collective excitations of
the D0-brane are described by the worldsheet embedding fields $x^i(z,\bar
z):\Sigma\to\real^9$, $i=1,\dots,9$, obeying Dirichlet boundary conditions on
$\del\Sigma$ while the target space temporal coordinate $x^0(z,\bar z)$
satisfies the standard Neumann boundary conditions\footnote{In the following we
will use Latin letters to denote spatial indices and Greek letters for
spacetime indices in the target space, which we assume for simplicity is a flat
Minkowski spacetime.}. The spatial embedding fields of $\Sigma$ are therefore
constant on its boundary while the temporal fields vary along $\del\Sigma$,
\beq
x^i(z,\bar z)\bigr|_{\del\Sigma}=y^i(x^0)~~~,~~~i=1,\dots,9
{}~~~~~~;~~~~~~\del_\perp x^0(z,\bar z)\bigr|_{\del\Sigma}=0
\label{bdryconds1}\eeq
where $y^i(x^0)$ are the collective coordinates of the D0-brane and $\del_\perp
$ denotes the normal derivative to the boundary of $\Sigma$. So the surface
$\Sigma$ is effectively a sphere with a marked point on it since the field $x$
is constant on the boundary. This means that an open string with both of its
ends resting on a single 0-brane is effectively a closed string forced to pass
through a fixed point in spacetime.

The action for the string propagation is defined by the $\sigma$-model
\beq
S_1=\frac1{4\pi\alpha'}\int_\Sigma \eta_{\mu\nu}\del x^\mu\bar\del x^\nu +
\oint_{\del\Sigma}y^i(x^0)\del_\perp x^i
\label{action1}\eeq
where $\eta_{\mu\nu}$ is a (critical) flat Minkowski spacetime metric. The
worldsheet boundary operator in \eqn{action1} describes the excitation of the
D-particle and it exploits the fact \cite{dlp,leigh} that the D-brane
configurations $y^i$ couple to the boundary vertex operator $\del_\perp
x^i|_{\del\Sigma}$. In this action formalism, appropriate for the computation
of vacuum amplitudes that we shall present, the Dirichlet boundary conditions
are $x^i(z,\bar z)|_{\del\Sigma}=0$ so that the boundary couplings $y^i$ depend
only on $x^0$. Using the alternative approach of implementing the non-zero
boundary conditions in \eqn{bdryconds1} would set the $y^i$-coupling in
\eqn{action1} to 0 \cite{leigh}. The two approaches are related in a simple
fashion at tree-level
of the semi-classical expansion. With the boundary conditions \eqn{bdryconds1},
the shift $x^i\to x^i+y^i$ sets $x^i|_{\del\Sigma}=0$ and the leading order
expansion of the bulk part of the action \eqn{action1} induces the boundary
term there. In general the transformation is more complicated and involves a
redefinition of the background fields \cite{leigh}. For the time being, we
suppress, for ease of notation, the explicit boundary operator in \eqn{action1}
which corresponds to the Neumann boundary condition on $x^0$ in
\eqn{bdryconds1}.

For a heavy non-relativistic D-particle we can write
\beq
y^i(x^0)=y^i(0)+u^ix^0
\label{Dpartvel}\eeq
where $u^i$ is the constant velocity of the D-brane. Upon T-dualization (see
section 4), the configurations $y^i(x^0)=-2\pi\alpha'A^i(x^0)$ form the
components of a dimensionally-reduced background $U(1)$ gauge field $A$, and
integrating out the embedding fields $x^\mu$ gives, to lowest order in the
derivative expansion, the Born-Infeld effective action \cite{ft,bachas}
\beq
\Gamma_{\rm BI}=\frac1{g_s}\int dt~\sqrt{1-\left(2\pi\alpha'F^{0i}\right)^2}=
\frac1{g_s}\int dt~\sqrt{1-(\dot y^i)^2}
\label{BIaction}\eeq
with $t$ the constant mode of the temporal embedding field $x^0$. The action
\eqn{BIaction} is just the relativistic free particle action for the D0-brane.
The beta-function equations for the renormalization group flow associated with
the worldsheet ultraviolet scale can be interpreted as the classical equations
of motion of the D-brane \cite{leigh}. The quantization of \eqn{BIaction} is
obtained by summing over all worldsheet topologies in the pinched approximation
as described in \cite{lm}. In the following we shall derive the generalization
of \eqn{BIaction} to the case of a multi D-particle system.

In the presence of two D-particles at positions $y^{(1)i}$ and $y^{(2)i}$ there
are excitations of the strings whose two endpoints are connected to a
single D-brane. These degrees of freedom are described by the fields
$x^{(a)\mu}$, with $a=1,2$, defined on two (identical) worldsheets
$\Sigma^{(a)}$. In addition there are excitations corresponding to strings
stretching between the two D-branes, which are described by a field
$x^{(12)\mu}$ defined on a worldsheet which is an annulus $\Sigma^{(12)}$. The
fields $x^{(12)i}$ have constant values $y^{(1)i}$ and $y^{(2)i}$ on the two
boundaries $\del\Sigma^{(12)}_1$ and $\del\Sigma^{(12)}_2$ of $\Sigma^{(12)}$,
respectively. The worldsheet $\sigma$-model action is a generalization of
\eqn{action1}
\beq\new{\begin{array}{lll}
S_{12}&=&\frac1{4\pi\alpha'}\sum_{a=1,2}\int_{\Sigma^{(a)}}
\eta_{\mu\nu}\del x^{(a)\mu}\bar\del
x^{(a)\nu}+\frac1{4\pi\alpha'}\int_{\Sigma^{(12)}}\eta_{\mu\nu}\del
x^{(12)\mu}\bar\del x^{(12)\nu}\\&
&~~~+\sum_{a=1,2}\left(\oint_{\del\Sigma^{(a)}}y_i^{(a)}(x^0)\del_\perp
x^{(a)i}+\oint_{\del\Sigma^{(12)}_a}y_i^{(a)}(x^0)\del_\perp
x^{(12)i}\right)\end{array}}
\label{action2}\eeq

Let us now consider what happens when the two D-branes coincide, i.e.
$y\(1=y\(2$. In this case the values of the field $x\({12}$ on the two
boundaries of $\Sigma^{(12)}$ coincide,
\beq
x\({12}\bigr|_{\del\Sigma^{(12)}_1}=\zeta x\({12}\bigr|_{\del\Sigma^{(12)}_2}
\label{bdryconds12}\eeq
where $\zeta=+1$ for unoriented and $\zeta=-1$ for oriented open strings. It is
then possible to express \eqn{action2} as the action for the collective
excitations of a single brane but with fields which are {\em matrix} valued. We
define the $2\times2$ symmetric matrix field
\beq
X^\mu=\pmatrix{x^{(1)\mu}& x^{(12)\mu}\cr x^{(12)\mu} & x^{(2)\mu}\cr}
\label{Xmatrix12}\eeq
The action \eqn{action2} then becomes formally equivalent to \eqn{action1},
\beq
S_{12}=\frac1{4\pi\alpha'}\int_\Sigma\tr~\eta_{\mu\nu}\del X^\mu\bar\del
X^\nu+\oint_{\del\Sigma}\tr~Y^i(x^0)\del_\perp X^i
\label{actionmat}\eeq
where the $2\times2$ matrix $Y^i$ is the boundary value of the matrix $X^i$,
\beq
Y^i(x^0)=y^i(x^0)\pmatrix{1&1\cr\zeta&1\cr}
\label{Ymatrix12}\eeq
and the worldsheet $\Sigma$ is a disc (equivalently a sphere with a single
marked point).

Thus the above worldsheet $\sigma$-model point of view leads to a rather simple
interpretation of the fact that, in the presence of coincident D0-branes,
spacetime becomes matrix-valued. These arguments generalize in the obvious way
to an arbitrary number $N$ of coincident D0-branes. The worldsheet
$\sigma$-model action is constructed from $N$ embedding fields $x^{(a)i}$,
$a=1,\dots,N$, describing the the collective excitations of the $N$ D0-branes
themselves, and from $\frac12N(N-1)$ fields $x^{(ab)i}$, with $1\leq a<b\leq
N$, describing the fundamental strings exchanged between the various D0-branes.
The matrix action \eqn{actionmat} is then defined in terms of $N\times N$
symmetric (for $\zeta=+1$) or antisymmetric (for $\zeta=-1$) matrices. It
possesses a global $O(N)$ symmetry defined by simultaneous rotations of the
matrices $X^i$ and $Y^i$.

\newsection{Quantum Dynamics of D-branes}

The mutually commuting matrix fields \eqn{Ymatrix12} describe strings with
their ends rigidly attached at the same point in space (coincident D-branes).
To describe the collective quantum dynamics of the system of D-branes we must
perturb $y_i^{(1)}=y_i^{(2)}+\varepsilon$ by a parameter $\varepsilon\to0$
which provides an infinitesimal separation between the D-particles. This
configuration arises, as we now demonstrate, from quantum fluctuations of the
system around the fixed configurations above, due to, for instance, recoil
effects from the scattering of closed string states off the D-brane background
\cite{lm}--\cite{emn} or from the emission of closed or open string states
representing interactions between the membranes. Likewise, the endpoints of the
fundamental string exchanged between the two branes should be considered to
differ from $y^{(a)i}$ by infinitesimal amounts, but to maintain the closed
nature \eqn{bdryconds12} only one more degree of freedom is added to describe
the boundary of the exchanged propagating string.

We consider the propagation of an open string in a single D-brane background,
corresponding, for example, to the collective position coordinates $y_i^{(1)}$.
In such a process the recoil of the D-brane background due to scattering by the
open string can be described by operators of the form~\cite{recoil}
\be
C(x^{(1)};y^{(1)})\equiv\epsilon y_i^{(1)}(0)\Theta (x^0)\partial_\perp
x^{(1)i}~~~~~~,~~~~~~D(x^{(1)};y^{(1)})\equiv
u_i^{(1)}x^0\Theta(x^0)\partial_\perp x^{(1)i}
\label{recoiloper}\ee
where $u_i^{(1)}$ is the recoil velocity of the brane, due to the string
emission or scattering, and the infinitesimal parameter $\epsilon$ regularizes
the step function $\Theta(x^0)$. It can be shown~\cite{recoil} that
$\epsilon^2$ is related to logarithmic divergences on the worldsheet via
$\log\Lambda\sim\epsilon^{-2}$, with $\Lambda$ the (ultraviolet)
renormalization group scale on $\Sigma^{(1)}$.

The sum over pinched worldsheet genera, which are the dominant configurations
in the random surface sum, leads to quantum fluctuations of the $\sigma$-model
couplings~\cite{emn}, in analogy with higher-dimensional wormhole
calculus~\cite{coleman}. Consider the partition function of an open string
$\sigma$-model in a recoiling D-brane background
\beq
Z^{(1)}=\sum_g\int[dx^{(1)}]~\e^{-S_0[x^{(1)}]-\oint_{\partial\Sigma^{(1)}}
C(x^{(1)};y^{(1)})-\oint_{\partial\Sigma^{(1)}}D(x^{(1)};y^{(1)})}
\label{pi}\eeq
where $S_0$ is the free $\sigma$-model action and the sum is over all
worldsheet genera $g$. In the case of weakly-coupled strings the genus
expansion can be truncated to a resummation over annulus graphs. Then the
dominant configurations in the sum are pinched worldsheet annuli, with
vanishing pinching size $\delta\to0$. Such insertions $\Delta S$ in the path
integral (\ref{pi}) are expressed in terms of bilocal operators ${\cal
O}_i{\cal O}_j$ on $\Sigma^{(1)}$, and exponentiate to give
\beq
{\sum_g}^{({\rm p})}\,\Delta S_g=\e^{\Delta
S}=\e^{\int\!\!\!\int_{\Sigma^{(1)}}{\cal O}_i{\cal
O}_j\delta^{ij}}=\int\prod_kd\alpha_k~
\e^{-\frac12\alpha_i\alpha_jg^{ij}}~\e^{\alpha^i\int_{\Sigma^{(1)}}{\cal O}_i}
\label{pi2}\eeq
where $\alpha_j$ are worldsheet wormhole parameters~\cite{coleman}, and
$g^{ij}$ is an appropriate metric in parameter space. The Gaussian weight in
(\ref{pi2}) can be thought of as a wormhole parameter space distribution
function. When (\ref{pi2}) is inserted into the path integral (\ref{pi}) one
obtains a quantization~\cite{emn} of the $\sigma$-model couplings $g^i$ of the
deformation operators $g^i\int_{\Sigma^{(1)}}{\cal O}_i$,
\beq
g^i\to\hat g^i=g^i+\alpha^i
\label{qcoupl}\eeq
and the $\sigma$-model coupling constants thus become operators in the target
space.

In the case of the boundary conformal field theory for the recoil operators
(\ref{recoiloper}), these operators carry {\it zero} conformal
dimension~\cite{recoil2} which leads to $\log\delta$ (modular) divergences in
the string propagator across thin worldsheet strips. The analysis of
divergences made in \cite{lm} reveals that the leading $\log\delta$ divergences
can be cancelled by momentum conservation, while subleading divergences can be
absorbed by renormalizing the widths of the wormhole distribution functions for
the $y_i$ and $u_i$ couplings
\beq
{\cal P}_{u_i}=\e^{-({\hat u}_i^{(1)}-u_i^{(1)})^2/2(\Delta
u^{(1)}_i)^2}~~~~~~,~~~~~~{\cal P}_{y_i}=\e^{-({\hat
y}_i^{(1)}-y_i^{(1)})^2/2(\Delta y^{(1)}_i)^2}
\label{distr}\eeq
The widths of the distributions are the respective quantum
uncertainties~\cite{lm}
\beq
(\Delta u_i^{(1)})^2\sim g_s^2\epsilon^2\log\delta~~~~~~,~~~~~~(\Delta
y_i^{(1)})^2\sim g_s^2\epsilon^{-2}\log\delta
\label{uncert}\eeq
One can absorb the $\delta\to0$ infinities into a renormalization
$g_s^2=(g_s^{\rm ren})^2/\log\delta$ of the string coupling constant, thereby
obtaining an uncertainty in the collective coordinates of the D-brane of order
$(\Delta y_i^{(1)})^2\sim(g_s^{\rm ren})^2\epsilon^{-2}$. For heavy branes,
their BPS mass $M_D\sim1/g_s^{\rm ren}\to\infty$ and one may assume that the
string coupling is weak enough so that $g_s^{\rm ren}\ll\epsilon^2$ implying
\beq
\Delta y_i^{(1)}\simeq\varepsilon\equiv g_s^{\rm ren}\epsilon^{-1}\ll1
\label{gstringappear}\eeq

Such infinitesimal fluctuations imply that
$y_i^{(1)}-y_i^{(2)}\sim\varepsilon$. This quantum uncertainty destroys the
classical rigidity of the two overlapping branes, and promotes the distance
between two quantum D-branes to a dynamical degree of freedom. This property
can be extended to strongly-coupled strings (for which the above perturbative
computation of the uncertainties is not strictly valid) by invoking a
strong-weak coupling duality. Thus to properly account for the sum over
worldsheet topologies the boundary matrix is represented in the generic
symmetric or antisymmetric form
\beq
Y^i=\pmatrix{y^{(1)i}&y^{(12)i}\cr\zeta y^{(12)i}&y^{(2)i}\cr}
\label{Ymatrix12dyn}\eeq

\newsection{T-dual Formalism}

The discussion of the previous section shows that to describe the quantum
dynamics of the multi D-brane system one needs to ``glue" two discs
$\Sigma^{(a)}$ representing the D-particles to the ends of the
``worldtube" $\Sigma^{(12)}$ which is a cylinder representing the fundamental
string between them. Alternatively, we glue a handle $\Sigma^{(12)}$ to the
surface of a sphere with two marked points. We must then allow the two ends of
the tube to fluctuate in spacetime. This can be described within a T-dual
formalism \cite{dornotto} for the Dirichlet action in \eqn{action2} for the
exchanged string with worldsheet cylinder $\Sigma^{(12)}$,
\beq
S_D[x^{(12)}]=\frac1{4\pi\alpha'}\int_{\Sigma^{(12)}}\eta_{\mu\nu}\del
x^{(12)\mu}\bar\del
x^{(12)\nu}+\sum_{a=1,2}\oint_{\del\Sigma_a^{(12)}}y_a^{(12)i}(x^0)\del_\perp
x^{(12)i}
\label{Dopenaction}\eeq
For this, we consider the path integral
\beq\new{\begin{array}{lll}
Z^{(12)}&\equiv&\int[dx^{(12)\mu}]~\e^{-S_D[x^{(12)}]}\\&=&\int[dx^{(12)\mu}]
{}~[dh^i]~[d\lambda^i]~\exp\left\{-\frac
1{\pi\alpha'}\int_{\Sigma^{(12)}}\left(h_z^ih_{\bar z}^i+h_z^i\bar\del
x^{(12)i}-h_{\bar z}^i\del x^{(12)i}\right)\right.\\& &~~~~~~~~~~~~~~~~~~
\left.+\frac1{4\pi\alpha'}\int_{\Sigma^{(12)}}\del x^0\bar\del
x^0+i\sum_{a=1,2}\oint_{\del\Sigma^{(12)}_a}\lambda^i_a
\left(x^{(12)i}-y_a^{(12)i}\right)\right\}\end{array}}
\label{Z12def}\eeq
The fields $h_{z,\bar z}^i$ implement a functional Gaussian integral transform
of the spatial part of the kinetic term in \eqn{Dopenaction}. We have also
exploited the fact that the string propagation is free, apart from the
Dirichlet boundary conditions at its endpoints which are implemented by
Lagrange multiplier fields $\lambda_a^i$. Integrating by parts over the
worldsheet yields
\beq\new{\begin{array}{c}
Z^{(12)}=\int[dx^{(12)\mu}]~[dh^i]~[d\lambda^i]~\exp\left\{-\frac1
{\pi\alpha'}\int_{\Sigma^{(12)}}\left(-\mbox{$\frac14$}\,\del x^0\bar\del
x^0+h_z^ih_{\bar z}^i-x^{(12)i}\left[\bar\del h_z^i-\del h^i_{\bar
z}\right]\right)\right.\\
\left.+\sum_{a=1,2}\left(i\oint_{\del\Sigma^{(12)}_a}
\lambda_a^i\left(x^{(12)i}-y_a^{(12)i}\right)+\frac1{\pi\alpha'}
\int_0^1h_{s_a}^ix^{(12)i}~ds_a\right)\right\}\end{array}}
\label{Z12intparts}\eeq
where $s_a\in[0,1]$ parametrize the circles $\del\Sigma^{(12)}_a$.

Integrating out $x^{(12)i}$ in the interior and $\lambda^i$ on the boundary of
$\Sigma^{(12)}$ in \eqn{Z12intparts} gives
\beq\new{\begin{array}{l}
Z^{(12)}=\int[dx^{(12)0}]~[dh^i]~\delta[\bar\del h_z^i-\del h_{\bar
z}^i]\\~~~~~~~~~~\times\exp\left\{-\frac
1{\pi\alpha'}\int_{\Sigma^{(12)}}\left(-\mbox{$\frac14$}\,\del x^0\bar\del
x^0+h_z^ih_{\bar z}^i\right)+\frac1{\pi\alpha'}\sum_{a=1,2}
\int_0^1y_a^{(12)i}h_{s_a}^i~ds_a\right\}\end{array}}
\label{Z12final}\eeq
The delta-function constraint on $h^i$ in \eqn{Z12final} implies that its Hodge
decomposition over the worldsheet has the form (neglecting harmonic modes)
\beq
h_z^i(z,\bar z)=\mbox{$\frac12$}\,\del\tilde x^{(12)i}(z,\bar
z)+\bar\del\xi^i(z,\bar z)~~~~~~,~~~~~~h_{\bar z}^i(z,\bar
z)=\mbox{$\frac12$}\,\bar\del\tilde x^{(12)i}(z,\bar z)-\del\xi^i(z,\bar z)
\label{hHodgedecomp}\eeq
with $\nabla^2\xi^i=0$. Substituting \eqn{hHodgedecomp} into \eqn{Z12final}, we
see that the transverse part $\xi^i$ of $h^i$ only contributes boundary terms
to the path integral, yielding some overall constants (as do the harmonic modes
of $h^i$). The effective action in \eqn{Z12final} thus depends only on the
longitudinal part of \eqn{hHodgedecomp}, and defining $\tilde x^{(12)0}\equiv
x^{(12)0}$ it can be written as
\beq
S_N[\tilde
x^{(12)}]=\frac1{4\pi\alpha'}\int_{\Sigma^{(12)}}\eta_{\mu\nu}\del\tilde
x^{(12)\mu}\bar\del\tilde x^{(12)\nu}+\sum_{a=1,2}\int_0^1C^i_a(\tilde
x^0)\frac{\del\tilde x^{(12)i}}{\del s_a}~ds_a
\label{Nopenaction}\eeq
where
\beq
C^i_a(\tilde x^0)=-\mbox{$\frac1{2\pi\alpha'}$}\,y_a^{(12)i}(\tilde x^0)
\label{Ciadef}\eeq

The action \eqn{Nopenaction} is the worldsheet $\sigma$-model action for an
open string with the usual Neumann boundary conditions $\del_\perp\tilde
x^{(12)i}|_{\del\Sigma^{(12)}_a}=0$ at its two endpoints which allow the
fields $\tilde x^{(12)i}(z,\bar z)$ to vary along the boundary of the
worldsheet. Generally, the path integral manipulation above implements
T-duality as a functional canonical transformation of the open string theory
\cite{dornotto}, mapping the spatial embedding fields $x^i$ to their duals
$\tilde x^i$ and converting the Dirichlet boundary conditions to Neumann ones.
Note that the temporal direction is not T-dualized ($x^0=\tilde
x^0$).

This equivalent, T-dual picture of the multi D-brane system thus shows how to
allow the string endpoints to fluctuate by replacing the Dirichlet boundary
conditions with Neumann ones on $\del\Sigma$, and at the same time it relates
the D-brane field theory to a ten-dimensional gauge theory \cite{taylor,kss}
which describes the low-energy sector of open superstring theory. Now that the
endpoints of the string which stretches between the two D-particles are allowed
to propagate in spacetime, it is possible to ``twist" the boundary
condition \eqn{bdryconds12} describing coincident D-branes. For this, we
augment each $y^{(12)i}=|y^{(12)i}|~\e^{i\theta_i}$ to generic {\it complex}
valued fields, such that their conjugates $(y^{(12)i})^*$ measure the changes
of orientation between the two boundaries. Then
$|y^{(12)}|\sim|y^{(1)}-y^{(2)}|$
effectively measures the relative distance between the two D-particles and the
angular variables $2\theta_i\in(0,2\pi]$ measure their relative orientation
along each direction in spacetime. For the canonical boundary values
\eqn{bdryconds12}, we have $\theta_i=\pi$ for $\zeta=+1$ (oriented strings)
while $\theta_i=\frac\pi2$ for $\zeta=-1$ (unoriented strings). The diagonal
components of \eqn{Ymatrix12dyn} remain real-valued since they correspond to
definite D-particle positions. Thus we replace \eqn{Ymatrix12dyn} by the
$2\times2$ {\it Hermitian} matrix field
\beq
Y^i(x^0)=\pmatrix{y^{(1)i}(x^0)&y^{(12)i}(x^0)\cr
y^{(12)i}(x^0)^*&y^{(2)i}(x^0)\cr}
\label{Ymatrix12herm}\eeq
Thus, in the general case of $N$ D-branes, the effective action for the
$\sigma$-model couplings will possess a
generic global $U(N)$ symmetry $Y^i\to UY^iU^\dagger$ corresponding to
rotations of the D-branes relative to one another. This yields a dynamical
worldsheet realization of the fact that at very short distances, spacetime is
described by mutually noncommutative matrix fields with global $U(N)$ symmetry.

\newsection{Open Strings in Fat Brane Backgrounds}

The dynamics of the system described in the previous sections, as we have so
far set it up, do not take into account the interactions between the
constituent D-branes and fundamental strings. Promoting the string embedding
fields to matrices as described in section 2 and T-dualizing, we have the
$U(N)$-symmetric action
\beq\new{\begin{array}{lll}
S_N&=&\frac1{4\pi\alpha'}\int_\Sigma\tr~\eta_{\mu\nu}\partial\tilde
X^\mu\bar\partial\tilde
X^\nu-\frac1{2\pi\alpha'}\oint_{\partial\Sigma}\tr~Y^i(x^0)\,d\tilde
X^i(s)+\oint_{\partial\Sigma}\tr~A^0(x^0)\,d\tilde
X^0(s)\\&=&\sum_{a=1}^N\left(\frac1{4\pi\alpha'}\int_{{\cal
D}^2}\eta_{\mu\nu}\partial\tilde x^{(a)\mu}\bar\partial\tilde
x^{(a)\nu}-\frac1{2\pi\alpha'}\oint_{S^1}y^{(a)i}(x^0)\,d\tilde
x^{(a)i}(s)+\oint_{S^1}A_{aa}^0(x^0)\,dx^0(s)\right)\\&
&+\sum_{a<b}\left(\frac1{2\pi\alpha'}\int_{{\cal A}^2}\eta_{\mu\nu}
\partial(\tilde x^{(ab)\mu})^*\bar\partial\tilde
x^{(ab)\nu}-\frac1{\pi\alpha'}\oint_{S^1}{\rm Re}~y^{(ab)i}(x^0)^*\,d\tilde
x^{(ab)i}(s)\right.\\& &~~~~~~~~~~~~~~~\left.+2\oint_{S^1}{\rm
Re}~A^0_{ab}(x^0)^*\,dx^0(s)\right)\end{array}}
\label{trivaction}\eeq
where $\Sigma$ is a disc ${\cal D}^2$ with a set of marked points on it, ${\cal
A}^2$ is an annulus, and we have explicitly written the boundary term with
$A^0$ which parametrizes the original Neumann boundary conditions on the
temporal
components, $\partial_\perp x^0|_{\partial\Sigma}=0$. The action
\eqn{trivaction} is a sum of $N^2$ actions for the independent fields
$x^{(a)\mu},x^{(ab)\mu}$ which describes the assembly of D-brane configurations
and fundamental strings. It is therefore missing the information about the
interactions between the various D-particles via exchanges of the fundamental
strings. Equivalently, the relabelling of the worldsheets between the first and
second lines of \eqn{trivaction} is not quite correct, i.e. the matrix version
of the multi D-brane action (describing a single composite D-particle with
matrix-valued coordinates as in section 2) does not properly account for the
different worldsheet topologies. The proper incorporation of these topologies
in fact breaks the $U(N)$ invariance group of the matrix action.

\begin{figure}[htb]
\epsfxsize=2in
\centerline{\epsffile{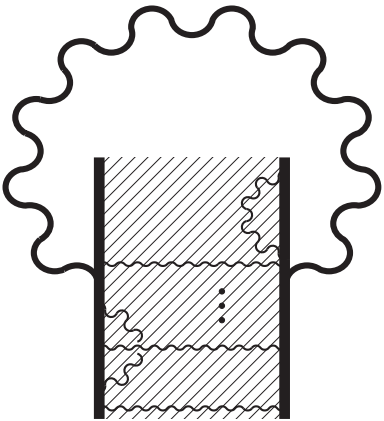}}
\caption{\it\baselineskip=12pt Schematic representation of a fat brane. The
bold strips denote the assembly of $N$ parallel D-branes and the thin wavy
lines represent the fundamental strings which start and end on them. The
shading represents the integration over all of these string interactions, as
well as the sum over genera described in section 3. The matrix $\sigma$-model
describes the interaction of the fat brane with a single fundamental string,
represented by the thick wavy line, which starts and ends on the fat brane with
a matrix-valued coupling constant $Y$.}
\label{fig.1}\end{figure}

We want to describe the $\sigma$-model of a background of $N$ interacting
branes, where the interactions are taken care of by open and closed strings
exchanged between them. For this, we shall use the composite configuration
described by \eqn{trivaction} to properly incorporate the dynamics of the
D-particles. We call this composite a ``fat brane" (Fig. \ref{fig.1}). Namely,
a fat brane is the above configuration of $N$ parallel D-branes with all string
interactions (including the sums over genera) among them integrated out. If we
consider the $\sigma$-model with worldsheet that has the topology of a disc,
then we integrate out the marked points associated with the various D-brane
boundary conditions $y^{(a)},y^{(ab)}$, leaving a low-energy effective theory.
This takes care of the topological configurations associated with the
worldsheet dynamics, at the cost of introducing matrix-valued couplings $Y^i$.
In the limit where the separations between the $N$ constituent D-branes becomes
large, the string interactions among them are negligible and the field theory
is the free $U(1)^N$ theory which is described by $N$ copies of the single
D-particle action discussed in section 2. When the relative separations vanish,
the branes fall on top of each other (giving a usual ``thin brane") and the
field theory is described by the composite D-brane $\sigma$-model of section 2
with matrix-valued fields and enhanced $U(N)$ symmetry. This presents an
alternative dynamical derivation of the short-distance noncommutativity of
spacetime, in which massless excitations in the spectrum of the quantum $U(N)$
gauge theory yield bound states of D-branes with broken supersymmetry (i.e.
$[Y^i,Y^j]\neq0$ for $i\neq j$) \cite{witten1}.

The effective matrix $\sigma$-model describes the target space (equivalently
coupling constant space) dynamics of a {\it single} fundamental string
propagating in the background of the fat brane. The string couples to the fat
brane background via a matrix-valued coupling constant $Y^i(x^0)$. By $U(N)$
invariance, the non-trivial matrix-valued interactions are described by the
expectation value, in a free $\sigma$-model, of the path-ordered $U(N)$ Wilson
loop operator $W[{\cal C}]$ along the boundary of the worldsheet,
\beq\new{\begin{array}{lll}
\sum_g\int[d\tilde X^\mu]~\int_\Sigma\prod_{a,b=1}^Nd^2z_{ab}~\e^{-S_N[\tilde
X;A]}&\simeq&\left\langle\!\left\langle W[{\cal
C}=\partial\Sigma]\right\rangle\!\right\rangle\\&\equiv&\int[d\tilde
x^\mu]~\e^{-N^2S_0[\tilde x]}~\tr~P\exp\left(i\oint_{\del\Sigma}A_\mu(\tilde
x^0(s))~d\tilde x^\mu(s)\right)\end{array}}
\label{Wilsonloop}\eeq
where $d\tilde X^\mu$ is the invariant Haar measure for integration over the
Lie algebra of $N\times N$ Hermitian matrices, $z_{ab}$ are marked points on
the disc $\Sigma$, and $A^\mu=(A^0,-\frac1{2\pi\alpha'}Y^i)$ can be interpreted
as a
ten-dimensional $U(N)$ isospin gauge field dimensionally reduced to the
worldline of the D-particle. The partition function \eqn{Wilsonloop} describes
the effective dynamics of an open string, with Neumann boundary conditions,
propagating in the fat brane background, the latter of which is described by
the matrix-valued coupling constants $Y^i(x^0)$ and incorporated by the
insertion of the Wilson loop in the free $\sigma$-model action of
\eqn{Wilsonloop}. The addition of this term can be thought of as either a
modification of the path integral measure which takes the boundary conditions
into account, or as the only non-trivial $U(N)$-invariant possible addition to
the action. Each matrix element $Y_{ab}$ represents the coupling of the
fundamental string to either a constituent D-brane (for $a=b$) or to an
exchanged string between the D-particles with a particular orientation (for
$a\neq b$).

Notice that the coordinates $\tilde x^i$ of the open string are scalars, as is
the common `time' $x^0$ for propagation of the fat brane background and the
open string. The time-dependence of $Y^i$ can be represented schematically as
\beq
Y_{ab}^i(x^0)=Y_{ab}^i(x^0=0)+U_{ab}^ix^0
\label{velocity}\eeq
in the simple case of non-relativistic heavy D-particles. Here $Y^i(x^0=0)$ is
the matrix-valued coupling constant (\ref{Ymatrix12herm}). The velocity matrix
$U^i$ describes the velocities of the constituent D-branes in the fat brane. We
can typically assume that the entire, composite fat brane moves with a single
(center of mass) velocity $u$, in which case $U_{ab}=u\delta_{ab}$.

The Neumann boundary conditions for $\tilde x^\mu$ in this case are
parametrized
by the quantities
\beq
C^\mu(\tilde x^0;s)=\bar\rho^a(s)A^\mu_{ab}(\tilde
x^0)\rho^b(s)
\label{Cigauge}\eeq
where $\bar\rho^a,\rho^b$ are Chan-Paton isospin factors at the endpoints of
the open string. The D-brane configurations can thus be identified with the
components of the gauge field $A_\mu$ restricted to the worldline of the
D-particle as
\beq
y^i(\tilde x^0)=-2\pi\alpha'\,\bar\rho^a(s)A^i_{ab}(\tilde x^0)\rho^b(s)
\label{Dgauge}\eeq
This relation can be used to map correlators of vertex operators of the open
string theory into expectation values of Wilson loop observables of the $U(N)$
gauge theory \cite{2b,kss,hamada}. For the non-abelian Chan-Paton factors
\eqn{Cigauge}, the worldsheet renormalization group equations have been derived
in \cite{dorn} and the (non-abelian) effective action in \cite{tseytlin1}. The
expression \eqn{Wilsonloop} no longer has the form of a functional integral
over the exponential of a local action. However, the path integral for the
Neumann $\sigma$-model action with the non-abelian Chan-Paton factors
\eqn{Cigauge} does \cite{dornotto}. Note that when $\bar\rho^a=\delta^{ac}$ and
$\rho^b=\delta^{bd}$, \eqn{Dgauge} reduces to the $\sigma$-model couplings
$Y_{ab}^i(\tilde x^0)$.

The T-dualization of this model using the background field method of the
previous section has been carried out in \cite{dornotto}, yielding the usual
D-brane $\sigma$-model action with a delta-function constraint for the boundary
conditions onto the abelian projection \eqn{Dgauge} of the gauge group. It is
crucial though to define the action in \eqn{Wilsonloop} using Neumann boundary
conditions in order to properly incorporate the coupling of the fundamental
string to the fat brane. The T-dual picture then shows that the resulting
$\sigma$-model action is in effect one for a {\it single} D-brane with abelian
projection coordinates \eqn{Dgauge}, as is anticipated from the analysis of
section 2. This is the essence behind the fat brane picture that we described
above and is the reason for the equivalence of the matrix $\sigma$-model action
with that for the collective quantum dynamics of multi D-brane configurations.

\newsection{Low-energy Effective Action}

The effective action for Wilson loop correlators of the sort \eqn{Wilsonloop}
has been derived recently by Tseytlin \cite{tseytlin1}, who showed that it
yields the natural non-abelian generalization of the Born-Infeld Lagrangian of
\cite{ft} for open strings in the background of non-abelian Chan-Paton factors.
To lowest order in the derivative expansion, the effective action depends only
on the field strength $F_{\mu\nu}=[D_\mu,D_\nu]$, with
$D_\mu=\partial_\mu-i[A_\mu,\,\cdot\,]$, and not on its gauge-covariant
derivatives. The result in the present case is then $\langle\!\langle
W[\partial\Sigma]\rangle\!\rangle\simeq\e^{-\Gamma_{\rm NBI}}$ with
\beq
\Gamma_{\rm NBI}=c_0\int dt~\tr~{\rm
Sym}\,\sqrt{\det_{\mu,\nu}\left[\eta_{\mu\nu}+2\pi\alpha'F_{\mu\nu}\right]}
\label{BI}\eeq
where $c_0$ is a constant, the trace is taken in the fundamental representation
of $U(N)$, and Sym denotes the symmetrized product ${\rm Sym}(A_1\cdots
A_n)=\frac{1}{n!}\sum_{\pi\in S_n}A_{\pi_1}\cdots A_{\pi_n}$.

The important technical point in the derivation of \eqn{BI} by the conventional
background field method is that covariant derivatives of the field strength
tensor $F_{\mu\nu}$ have been effectively set to zero, i.e. $D_\mu
F_{\nu\rho}=0$ \cite{tseytlin1}. In the case at hand the components of the
field strength are given by
\beq
2\pi\alpha'F_{0i}=D_0Y_i\equiv\dot
Y_i-i[A_0,Y_i]~~~~~~,~~~~~~(2\pi\alpha')^2F_{ij}=[Y_i,Y_j]
\label{fscomps}\eeq
since the coordinates $Y_{ab}$ of the fat brane do not depend on $x^i$, but
only on the Neumann time coordinate $x^0$ of the open string propagation in the
fat brane background. This implies that the condition for the vanishing of the
covariant derivative reads (in the gauge $A_0=0$)
\beq
2\pi\alpha'\partial_0F_{0i}=\ddot
Y_i=0~~~~~~,~~~~~~\partial_0[Y_i,Y_j]=0~~~~~~,~~~~~~[Y_j,[Y_i,Y_k]]=0
\label{covariantder}\eeq
For the simple boost (\ref{velocity}) in a non-accelerating fat brane
background, the conditions \eqn{covariantder} are indeed satisfied. Note also
that the non-trivial solitonic D-brane configurations correspond to
\cite{bfss,2b}
\beq
[Y_i,Y_j]=f_{ij}I_N
\label{bpsconfs}\eeq
where $I_N$ is the $N\times N$ identity matrix, and hence they also satisfy
\eqn{covariantder}.

The matrix model effective action \eqn{Matrixaction} now follows, after
appropriate rescalings of the fields, from the expansion of (\ref{BI}) in
powers of the Regge slope $\alpha'$. The linear terms vanish, while the first
non-zero term is the quadratic one which yields the bosonic part of
(\ref{Matrixaction}). Thus the (bosonic part of the) matrix model action can
be derived as the target space effective action of the non-local $\sigma$-model
(\ref{Wilsonloop}) describing the motion of an open string in a fat D-brane
background. The crucial aspect of this fat brane description is the
matrix-valued couplings $Y^i$ of the $\sigma$-model. For non-diagonal coupling
matrices, the coefficient of the commutator term $[Y^i,Y^j]^2$ is determined
dynamically from the quantum uncertainties \eqn{gstringappear} arising from the
quantum dynamics of the multi D-brane system which effectively renders the
matrices $Y^i$ mutually noncommutative. This picture therefore yields a
geometrical origin for the appearance of the string coupling constant $g_s$
(here in its renormalized form) in the matrix model action\footnote{An explicit
appearence of $g_s$ also occurs in the scattering of strings off a single
fluctuating quantum D-brane background \cite{emn}.}. Furthermore, these
same uncertainties naturally determine the solitonic BPS configurations
\eqn{bpsconfs} from a quantum mechanical origin.

We have shown that the matrix model for M Theory describes the propagation of
open strings in a background of $N$ interacting D-branes, and a worldsheet
$\sigma$-model description of this process is a (low-energy) effective theory
in which the string interactions of the multiple D-brane
configurations have been integrated out. In this sense, the open string
propagates in the background of a fat brane with a matrix-valued coupling
constant (\ref{Ymatrix12herm}). One important aspect that is not clear from
this construction is how to implement target space supersymmetry at the level
of the worldsheet $\sigma$-model. A supersymmetric extension of the abelian
Born-Infeld effective Lagrangian for a single D-brane has been derived recently
in \cite{schwarz}, although the $\sigma$-model origins of this construction are
not evident. One could mimick the fat brane construction of section 5 by
replacing \eqn{Wilsonloop} with a supersymmetric Wilson loop correlator in the
Green-Schwarz light-cone formalism \cite{hamada}. However, the T-dualization of
the resulting $\sigma$-model action to one with Dirichlet boundary conditions
appears to be highly non-trivial, even in the single D-particle case.

\end{document}